\documentclass[conference,10pt]{IEEEtran}

\usepackage{ifpdf}

\usepackage{cite}

\newif\ifarxiv
\arxivtrue

\ifCLASSINFOpdf
  \usepackage[pdftex]{graphicx}
  \ifarxiv
    \graphicspath{{./}}
  \else
    \graphicspath{{media/}}
  \fi
\else
  \usepackage[dvips]{graphicx}
  \graphicspath{{media/}}
\fi

\usepackage{amsmath}

\usepackage{array}

\ifCLASSOPTIONcompsoc
  \usepackage[caption=false,font=normalsize,labelfont=sf,textfont=sf]{subfig}
\else
  \usepackage[caption=false,font=footnotesize]{subfig}
\fi

\ifarxiv
\else
\usepackage{dblfloatfix}
\usepackage{fixltx2e}
\fi

\ifarxiv
  \usepackage{hyperref}
\else
  \usepackage[draft]{hyperref}
\fi
\usepackage[capitalize]{cleveref}
\usepackage{url}

\usepackage[color=yellow!60,textsize=footnotesize,obeyDraft,draft]{todonotes}
\usepackage{enumitem}
\ifarxiv
\else
  \usepackage{tikz}
  \usetikzlibrary{switching-architectures}
  \usepackage{tikzscale}
\fi

\usepackage[detect-weight=true, binary-units=true, per-mode=symbol, per-symbol=/]{siunitx}
\usepackage[all]{nowidow}
\usepackage{enumitem}
\usepackage{amssymb}
\usepackage{algorithm}
\usepackage{algpseudocode}

\usepackage{listings}
\usepackage{float}

\floatname{algorithm}{Listing}

\makeatletter
\newcommand{\crefnames}[3]{%
  \@for\next:=#1\do{%
    \expandafter\crefname\expandafter{\next}{#2}{#3}%
  }%
}
\makeatother
\crefnames{algorithm}{Listing}{Listing}

\newcommand{\sgnl}[2]{$\texttt{#1}_{#2}$}

\begin{document}

\bstctlcite{IEEEexample:BSTcontrol}

\title{A Bene\u{s} Based NoC Switching Architecture for Mixed Criticality
Embedded Systems}

\author{\IEEEauthorblockN{Steve Kerrison, David May and Kerstin Eder}
\IEEEauthorblockA{University of Bristol, Department of Computer Science\\
  Merchant Venturers Building, Woodland Road\\
  Bristol, BS8 1UB, United Kingdom.\\
Email: \{firstname.lastname\}@bristol.ac.uk}}
\date{April 2016}

\maketitle
\begin{abstract}

Multi-core, Mixed Criticality Embedded (MCE) real-time systems require
high timing precision and predictability to guarantee there will be no
interference between tasks. These guarantees are necessary in application areas
such as avionics and automotive, where task interference or missed deadlines
could be catastrophic, and safety requirements are strict. In modern multi-core
systems, the interconnect becomes a potential point of uncertainty, introducing
major challenges in proving behaviour is always within specified constraints,
limiting the means of growing system performance to add more tasks, or provide
more computational resources to existing tasks.

We present MCENoC, a Network-on-Chip (NoC) switching architecture that provides
innovations to overcome this with predictable, formally verifiable timing
behaviour that is consistent across the whole NoC. We show how the fundamental
properties of Bene\u{s} networks benefit MCE applications and meet our
architecture requirements. Using SystemVerilog Assertions (SVA), formal
properties are defined that aid the refinement of the specification of the
design as well as enabling the implementation to be exhaustively formally
verified. We demonstrate the performance of the design in terms of size,
throughput and predictability, and discuss the application level considerations
needed to exploit this architecture.

\end{abstract}

\IEEEpeerreviewmaketitle

\section{Introduction}

Real-time embedded systems perform a broad range of processing tasks, many of
which must take place within hard deadlines in order to avoid loss of
functionality or risks to safety. Modern embedded systems are prolific in many
safety-critical areas, including automotive, industrial and avionics. Alongside
safety, these sectors must deliver increasingly sophisticated features, such as
visual processing and automation, requiring increases in computational
capabilities and task count, without compromising safety.

This feature growth has necessitated multi-core solutions in-line with
processor scaling trends\cite{itrs2013}. However, the need to preserve
hard-real time and safety requirements in critical tasks creates a unique
problem set that must be handled at all levels, from architecture, through the
software development process and up to certification. This is further
confounded in large systems, where tasks with differing levels of criticality
must co-exist \cite{mcavionics2012}.

This paper addresses the interconnect challenges of multi-core, MCE systems. We
propose a NoC architecture that is designed to be both highly predictable in
its routing and timing behaviour, as well as formally verifiable. In doing so,
the architecture provides strong behavioural guarantees that allow
mixed-criticality tasks to be scheduled into a multi-core system. This aligns
with efforts to provide determinism at higher levels, such as Time Triggered
Ethernet\cite{tterationale} for distributed systems. We show that, despite
aggressive timing precision requirements, the architecture is scalable and
continues to be verifiable due to the simple structures from which it is built.
The main contributions of this work are:

\begin{itemize}

  \item Specification of a NoC that meets multi-core MCE needs.

  \item A novel, non-blocking and timing-predictable implementation addressing
    these requirements.

  \item Demonstration of scaling properties and performance.

  \item Formal verification to prove correct behaviour.

\end{itemize}

This paper is structured as follows. In \cref{sec:related} we discuss related
work. \Cref{sec:requirements} defines the requirements for the switching
architecture. The implementation is explained in \cref{sec:architecture}, then
its performance evaluated in \cref{sec:evaluation}. Software-level scheduling
considerations are discussed in \cref{sec:scheduling}. Formal verification is
presented in \cref{sec:formal}, including proof scaling results. Finally,
\cref{sec:conclusions} states our conclusions and proposes future work.

\section{Related work}
\label{sec:related}

Addressing mixed-criticality multi-core communication builds upon several areas
of research: multi-core systems; mixed criticality hardware and software along
with its certification challenges; and network architectures, in particular
NoC, to facilitate communication between large numbers of nodes in a single
device. 

\subsection{Multi-core MCE systems}

Traditional multi-tasking requires time-slicing of tasks onto a single
processor. However, as processor operating frequencies tend to no longer
increase, further performance is now more readily achieved through the addition
of more processors\cite{itrs2013}. The exchange of data between tasks is
typically achieved using a shared-memory model, where locations in memory are
accessed by multiple tasks. Alternatively, message passing may be used,
adopting a model such as CSP-style communication. In either case, the movement
of data between processing elements and memory has increasing complexity to
ensure high performance and data consistency, although there are claims that
this can continue to scale\cite{cacheCoherency}.

In a critical systems context, it is insufficient to provide increased
performance through multi-core. Predictability must be preserved along with
other protective measures to ensure deadlines are met and tasks do not
interfere. For example, adding a task to an otherwise unused core in a system
may not intuitively affect other tasks, but its network and memory access
patterns may in fact do so. To guarantee safety, it must therefore be proven
that this is not the case. Minimizing the effort required to do this, by
ensuring the underlying architecture can provide the necessary behavioural
guarantees, is then clearly desirable.

\subsection{Mixed criticality hardware and software}

A critical task has some function that either cannot be interrupted or has a
hard deadline to maintain. In a less critical task, interruption to service or
a deadline miss may be less important. For example, in an automotive context,
the braking system must respond to any sensor reading that indicates a loss of
traction under braking. If the response is delayed, or sensor data missed,
safety of the vehicle may be negatively affected. However, an in-car
entertainment system can tolerate occasionally dropped or corrupted video
frames, to within some defined Quality-of-Service (QoS), and safety may be less
of a concern.

Despite the clear differences in criticality in the above examples, there may
be some interaction between the tasks, as the entertainment system may be
integrated with the driver's controls and visual feedback. Therefore, these
activities are not necessarily completely isolated, but the most critical task
must clearly not be negatively affected by behaviour of the less critical one.
This presents a challenge in meeting certification requirements, of which there
may be several, whilst ensuring that resource allocation and scheduling allows
all tasks to operate correctly in the broadest set of
conditions\cite{mceCert2010}.

Hardware must support mixed-criticality, first by providing sufficient
resources that guarantee hard real-time behaviour. In this case, predictable
timing is essential. Real-time processors often sacrifice performance
enhancements such as cache-hierarchies in order to deliver such predictability.
Secondly, hardware must prevent tasks from interfering with each other, for
example by writing to incorrect memory regions or by creating resource
starvation.

The software must be suitably predictable to ensure that underlying hardware
guarantees can be provably exploited. Worst Case Execution Time
(WCET)\cite{DBLP:journals/tecs/WilhelmEEHTWBFHMMPPSS08} analysis may be
performed on a task to ensure that it completes within a desired deadline. In a
system of tasks, scheduling must take place, either online or offline.
Multi-core MCE systems pose additional problems over single-core solutions. In
a shared memory system, there may be contention for access to the memory
hierarchy or other interconnects\cite{mcavionics2012}. These may reduce
predictability and require more complex scheduling efforts or over-provisioning
of resources to guarantee safe behaviour. This exposes a need to provide more
predictability within the interconnect, to tighten these bounds, aiding
analysis and certification.

\subsection{NoC architectures}

A Network-on-Chip (NoC) is a collection of resources on a single chip,
typically including processing elements (cores), memory elements (caches,
DRAMs) and peripheral components (external interfaces, timers, etc). These are
all interconnected by one or more networks, departing from traditional bus
architectures in favour of a more scalable, routed arrangement. The intent of
NoC is to deliver a chip with processing elements numbering in the tens,
hundreds or more, thus achieving continued performance growth through many-core
scaling. In a multi-core MCE context, NoC is therefore highly desirable in
emerging chip designs, but one that must meet the constraints outlined earlier
in this section.

A variety of NoC topologies exist. Common approaches include mesh structures,
rings and higher-dimension extensions of these. The Xeon Phi\cite{XeonPhi2013}
uses a ring network to connect its 63 processing elements together along with
tag caches and memory controllers. Progress into multi-layer 3D stacked
processors~\cite{n3xt} is extending NoC requirements into three physical
dimensions.

Multiple mesh networks can be used to handle the different traffic patterns and
access requirements, for example physically separating network memory accesses,
core-to-core exchange and peripheral communication. Such approaches are seen in
the five-network TILE64\cite{Bell2008} and three-network Adapteva
Epiphany\cite{Adapteva}. Alternatively, the network may be segmented through
Time-Division Multiplexing (TDM), such as in the picoChip\cite{PicoChip}
processor, where bandwidth is guaranteed at pre-defined times, but may be
over-provisioned.

Due to the size of NoC systems, congestion handling and fault tolerance must
also be considered. Groups of cores may have more active communication, and
thus benefit from using additional nearby routes. Failures may require message
re-transmission~\cite{REPAIR}, re-routing, or be mitigated through redundant
hardware\cite{3dredundancy}.  Dynamic mechanisms for routing, particularly
unconstrained or non-deterministic, can be undesirable in a critical systems
context because they reduce the certainty of timing in message delivery.
Further, any network where messages may be blocked by other messages can
dramatically increase the upper bound for communication time, resulting in an
infeasible system specification or conditions that are unsafe for critical
tasks.

\subsection{Clos and Bene\u{s} networks}

Large scale interconnectivity has previously been addressed in
telecommunications, where circuits must be formed between telephone endpoints,
focused through potentially several telephone exchanges locally, nationally or
internationally. It was required to develop interconnects that did not place
infeasible wiring overheads or switch complexity upon the exchanges.
Clos\cite{Clos1952} described a multi-stage network of smaller crossbar
switches, layered in order to provide $N:N$ connectivity with lower wiring
complexity than a monolithic solution. Timing changes or re-routing artefacts
such as clicking are easily detected by the human ear, hence such networks must
minimize or completely remove such occurrences. This strictness bears many
similarities to the requirements of modern MCE systems.

\begin{figure}
  \vspace{-0.75em}
  \centering
  \subfloat[Clos\label{fig:clos}]{
    \includegraphics[width=0.425\columnwidth,clip,trim=8mm 55mm 120mm 8mm]{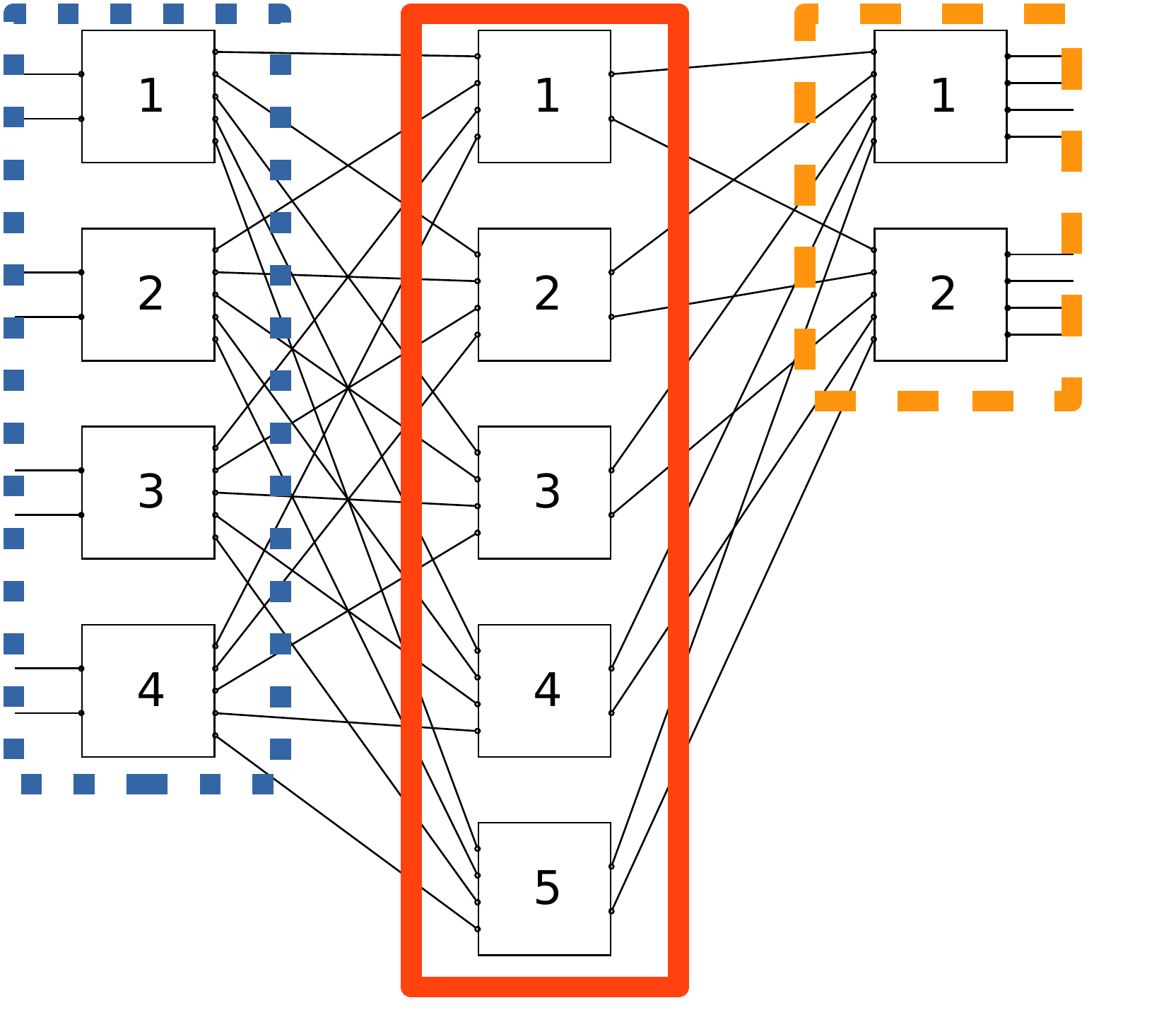}
  }
  \subfloat[Bene\u{s}\label{fig:benes}]{
    \includegraphics[width=0.45\columnwidth,clip,trim=8mm 55mm 110mm 8mm]{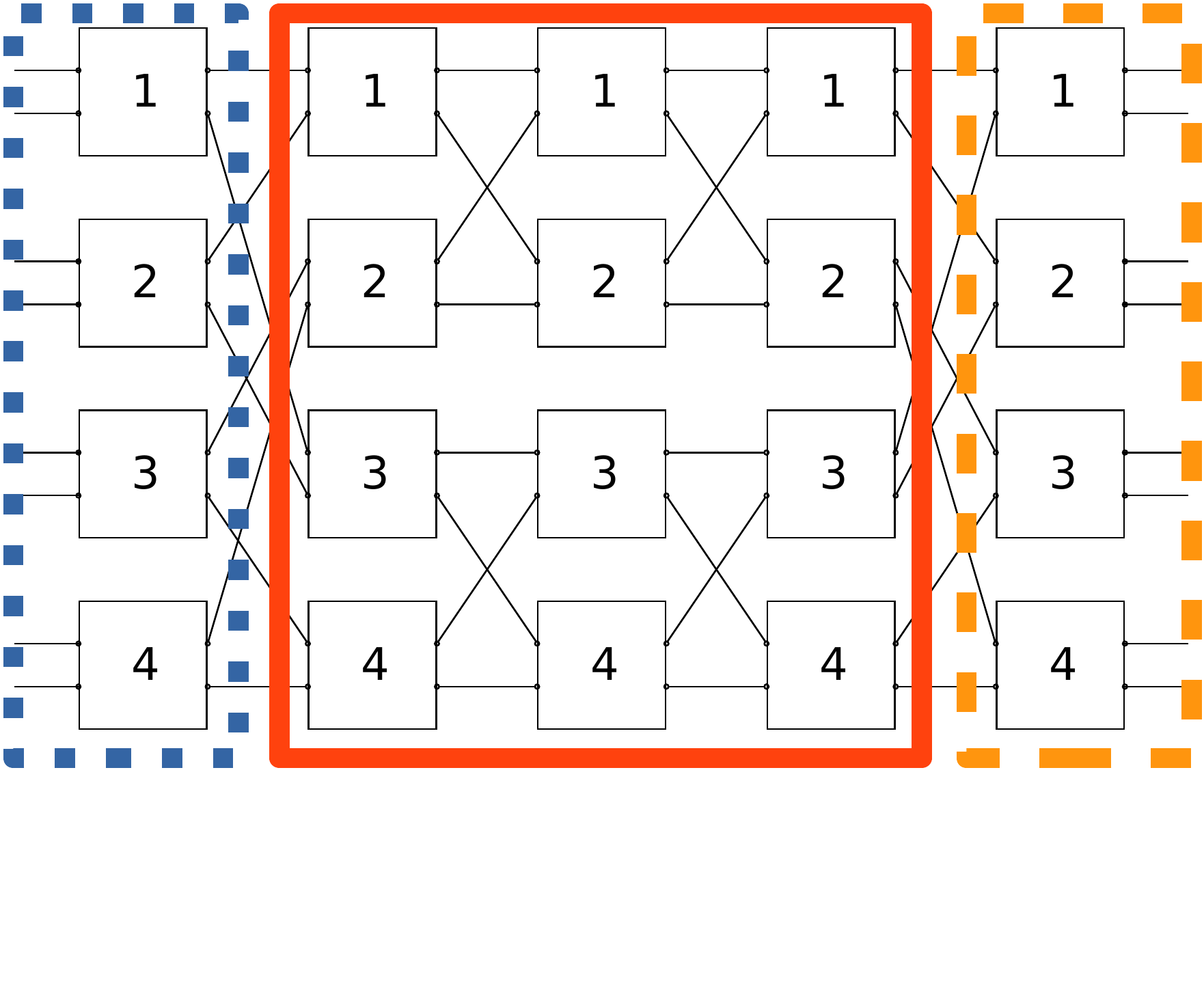}
  }
  \caption{Example of eight-port Clos and Bene\u{s} network structures with equivalent sections highlighted.}
  \label{fig:clos-benes}
\end{figure}

\Cref{fig:clos} depicts a Clos topology. It comprises three switching stages
where the degree of the switches different at each stage. A
Bene\u{s}\cite{Benes1962} network refines the Clos concept into a topology of
2x2 switching elements. It retains the same properties, but is constructed from
the smallest possible switch size, shown in \cref{fig:benes}. Any larger switch
in a Clos or Bene\u{s} network can be realised using sub-networks of Bene\u{s}
switches.

For our Mixed Critcality Embedded Network-on-Chip (MCENoC), particular
properties provided by these networks are desirable for several reasons:

\begin{enumerate}

  \item The number of routing stages for any communication is consistent.

  \item It is proven that any $N:N$ permutation of two-party communications is
    always routable in Bene\u{s} structures\cite{Benes1962, Pan2008}.

  \item The number of stages and switching elements used scales logarithmically
    with the number of nodes.

  \item The switching elements can be built upon simple crossbars, which are
    straightforward to create in VLSI\cite{jiang2014} and are then replicated
    many times.

\end{enumerate}

Many variations on these networks exist, as well as strategies for routing on
them\cite{Pan2008}. However, we focus on preserving predictability, whilst
keeping the core design simple and scalable, therefore the MCENoC design
choices reflect this.

\section{Requirements}
\label{sec:requirements}

In this section we define high-level requirements for the MCENoC design. These
are refined into a specification that guides the implementation. Both the
specification and implementation are subjected to scrutiny through the formal
verification process detailed in \cref{sec:formal}.

The aim of this work is to provide a communication architecture that enables
real-time embedded software to have precise behavioural guarantees in a
multi-core context. With this, the behaviour of the entire system can be
reasoned about statically, thus allowing verification and certification of
applications using the system. To focus this effort, several requirements are specified.

\paragraph*{All nodes in the network can be considered equidistant}

Predictability is simplified significantly when communication between nodes is
a uniform distance, as placement of computation and data is less constrained.

\paragraph*{Latency through the network is tightly bounded}

Both worst- and best-cases should be known to provide guarantees of expected
performance and satisfy task deadlines.

\paragraph*{Communication between two nodes is non-blocking}

Refining the previous requirement, by removing blocking, the flow of data
through the network is simpler to model, analyse and produce bounds for.
Buffering and flow control may still be present in network endpoints, however.

\paragraph*{Invalid or erroneous communications cannot interfere with more critical communications}

Where multiple flows exist within the network, those attached to critical tasks
must have priority in the network. Less critical tasks, which may have less
strict certification and verification requirements, must not be able to steal
routes required by higher criticality tasks, and a failure condition should be
asserted if this is attempted.

\paragraph*{The switching architecture should scale into many cores}

The target is a design that can ultimately scale up to hundreds of cores or
more. Thus, the logic utilisation and internal connectivity scaling should
support this.

\paragraph*{Routing decisions must be statically resolvable with minimal
overhead at large scales} All online control should be conducted in-band,
without a central arbiter of other similar global control mechanism. A large
amount of uncertainty can then be removed from the system's communication
profile.

\section{MCENoC architecture}
\label{sec:architecture}

\begin{figure}
  \centering
  \includegraphics[width=0.5\columnwidth,clip,trim=0.4cm 24.25cm 14.2cm 0.75cm]{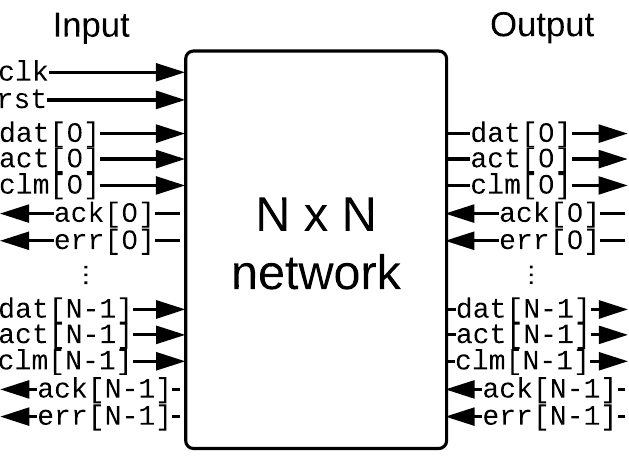}
  \caption{Top-level signalling of the design, valid as a per-switch or whole-network abstraction.}
  \label{fig:top}
\end{figure}

The MCENoC architecture is constructed from a
replicated set of relatively simple switches, forming a Bene\u{s} network. The
edges of the network then provide an interface to nodes, which can be
processing elements, memory, or other peripherals. Interface bridging is
expected to take place at he edge, with certain system level protocol
constraints imposed by the network. The design is implemented in SystemVerilog
2009 and includes a number of SystemVerilog Assertion (SVA)\cite{Mehta2014}
properties for validation of the specification and verification of the design.
In this section we first address the switch implementation, then the network
construction, followed by system level considerations.

\subsection{Switching element}

Each switching element is configurable to have $2^p$ ports, where $p$ is the
number of bits required to define a route through the switch. A block-level
diagram of signals is shown in \cref{fig:top}, with an arbitrary number of
ports with signals denoted as \sgnl{sig}{0} \dots \sgnl{sig}{N-1}. Each port has
both forward- and backward-propagated signals, allowing the transfer of data,
flow control and error condition between network stages, as well as permitting
an idle state. 

With no allocated ports, all outputs are set to their defaults and there is
scope for fine-grained clock gating. Allocation of a port $q$ is achieved by
clocking in $p$ configuration bits into \sgnl{dat}{q} whilst asserting the
\sgnl{clm}{q} and \sgnl{act}{q}. The configuration bits specify the target port
number, $r$. Following this, input $q$ is connected to output $r$ and
propagation of future signals takes place, through a buffer of depth one.

\subsubsection{Conflict identification and resolution}

An output $r$ can only be successfully allocated if no other input port is
already using it. If the requested output port is not in an idle state, then a
conflict is present, where the port is either already claimed, or multiple
inputs are simultaneously attempting to claim it. In the former case,
\sgnl{err}{q} is asserted until \sgnl{clm}{q} is de-asserted, after which a new
attempt at routing can begin. In the latter case, the input port with the
lowest value of $q$ obtains the route, with all other claims being rejected.

\subsubsection{Port states}

The switch input ports each have four possible states.

\begin{description}[font=\normalfont\itshape]

  \item[Wait:] {The input is not connected to any output and fewer than $p$
    configuration bits have been provided.}

  \item[Accept:] {The input is connected to the requested output and is
    propagating data.}

  \item[Reject:] {There were $p$ configuration bits received, but a conflict or
    protocol violation was observed and the port must be unclaimed before
  attempting configuration again. No output connection was made.}

  \item[Abort:] {A previously accepted connection is being destroyed due to an
    incoming error condition from the forward switching stage.}

\end{description}

If all input ports are in the \emph{wait} state and no \texttt{clm} signals are
asserted, the switch can be considered idle and may be clock-gated. Successful
connections can be destroyed either from the initiator, by de-asserting the
relevant \texttt{clm} signal, or from the destination, by asserting
\texttt{err}.

\subsection{Network construction}

To determine the network construction for an $N$ node network, a switching
element size, $B$, must be selected. If raising $B$ to a positive integer value
(in the set of natural numbers, $\mathbb{N}$) yields $N$, then the number of
stages is:
\begin{equation}
  S = 2\log_B(N) - 1, \quad \mathrm{if} \quad \exists x \in \mathbb{N} \mid B^x
  = N.
\end{equation}

Otherwise, any $N = 2^x$ port network, where $B \ge 2$, has a middle stage with
$m$ directional bits ($2^m$ ports per switch), where $2^m \le B$, such that:
\begin{subequations}
\begin{align}
  X &= \lceil \log_B(N) \rceil, \\
  S &= 2 \log_B(X) - 1,\\
  m &= \log_2({\textstyle \frac{N}{B^{X-1}}}).
\end{align}
\end{subequations}

\begin{figure}
  \vspace{-2em}
  \centering
  \subfloat[Four- and two-port switches]{
    \ifarxiv
      \centering
      \includegraphics[height=113px]{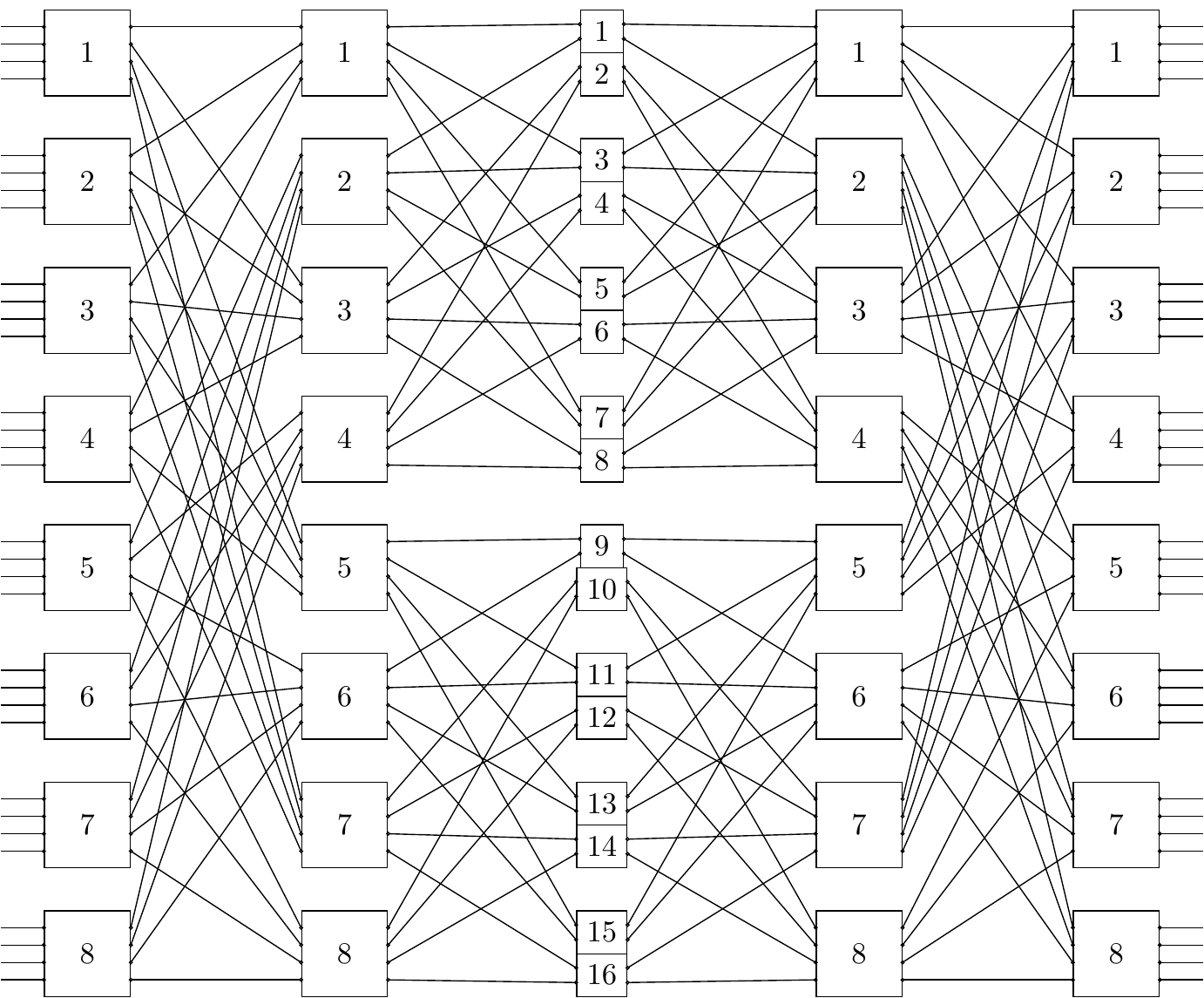}
    \else
      \begin{tikzpicture}[scale=0.35,every node/.style={scale=0.35}]
        \input{media/net-32port-2bit.tikz}
      \end{tikzpicture}
    \fi
  }
  \subfloat[Eight-port switches]{
    \ifarxiv
      \centering
      \includegraphics[height=123px]{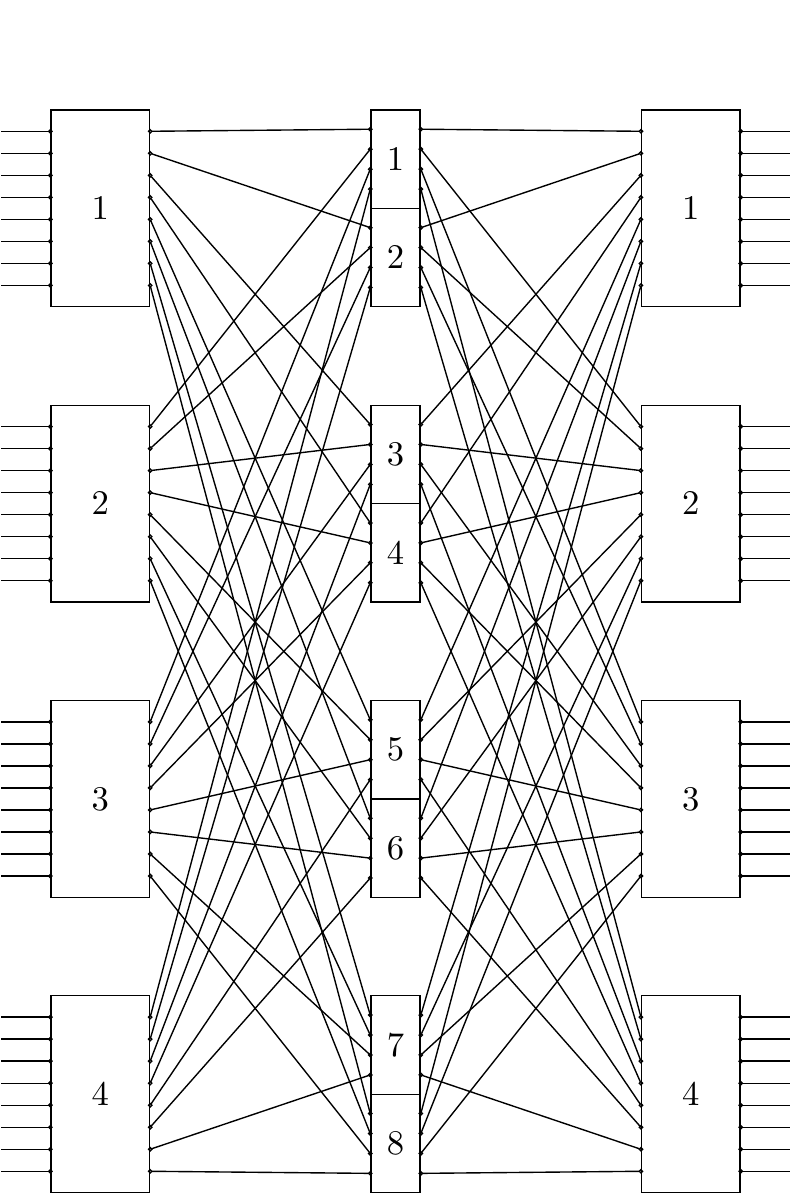}
    \else
      \begin{tikzpicture}[scale=0.37,every node/.style={scale=0.35}]
        \input{media/net-32port-3bit.tikz}
      \end{tikzpicture}
    \fi
  }
  \caption{Two 32-port networks using four- and two-port switches vs. eight- and four-port switches.}
  \label{fig:mcenoc-variants}
\end{figure}

For example, if the desired switch element size is four ports, then a 32-port
network will contain four 4-port stages and a middle stage of twice as many
2-port switches, whereas a 64-port network is simply five 2-port stages.  Two
32-port network examples are shown in \cref{fig:mcenoc-variants}. All network
diagrams are generated in TikZ that is emitted from SystemVerilog during the
\texttt{initial} phase of simulation, giving a direct representation of the
structure that is in use.

The connectivity of ports in the network, $c_n^i$, of index $i$ and at each
stage $n$ is shown in \cref{eq:connect}. Due to the symmetry of the network, it
can be considered as two halves connected in a similar progression, where $n =
0$ is the middle stage. The $i$th port in the inner stage connects to the $j$th
port in the next stage, wrapped in accordance to the block size at that stage.
The block size, $b_n$, is determined by the stage, $n$, where the total number
of connection stages, $s$, for half of the network, is obtained using $S$. The
connectivity formula accepts any power-of-two size of switching element, $B$.
\begin{subequations}
  \label{eq:connect}
  \begin{align}
    c_{n+1}^{j} \longleftrightarrow & c_{n}^{i}, \quad \mathrm{where:}\\
    j &= { \textstyle \left( k + \lfloor \frac{k}{b_n} \rfloor \right) \bmod b_n} + o, \\
    k & = (i - o) \cdot B, \quad o = { \textstyle \lfloor \frac{i}{b_n} \cdot b_n \rfloor }, \\
    b_n &= { \textstyle \min\left(B^{2+s-(s-n)}, N\right) }, \quad s = { \textstyle \frac{S-1}{2} }.
  \end{align}
\end{subequations}

The network can be considered folded horizontally, where input port $q$ and
output port $r$ belong to the same node.

The topology of the network guarantees that on an $s$ stage
network requiring $p$ routing bits, a route can be established $p + s$ cycles.
Following this, an input can propagate through the network in $p$ cycles. In
the case where a route conflict occurs, the worst-case latency in the input
receiving an \texttt{err} signal is $2p + s$, where the conflict occurs in the
final stage and the error must propagate back through all prior stages.

\subsection{System implementation}

\begin{figure*}
  \vspace{-1em}
  \centering
  \subfloat[Utilisation\label{fig:util}]{\includegraphics[width=0.32\textwidth,clip,trim=0.5cm 0.5cm 0.5cm 0.5cm]{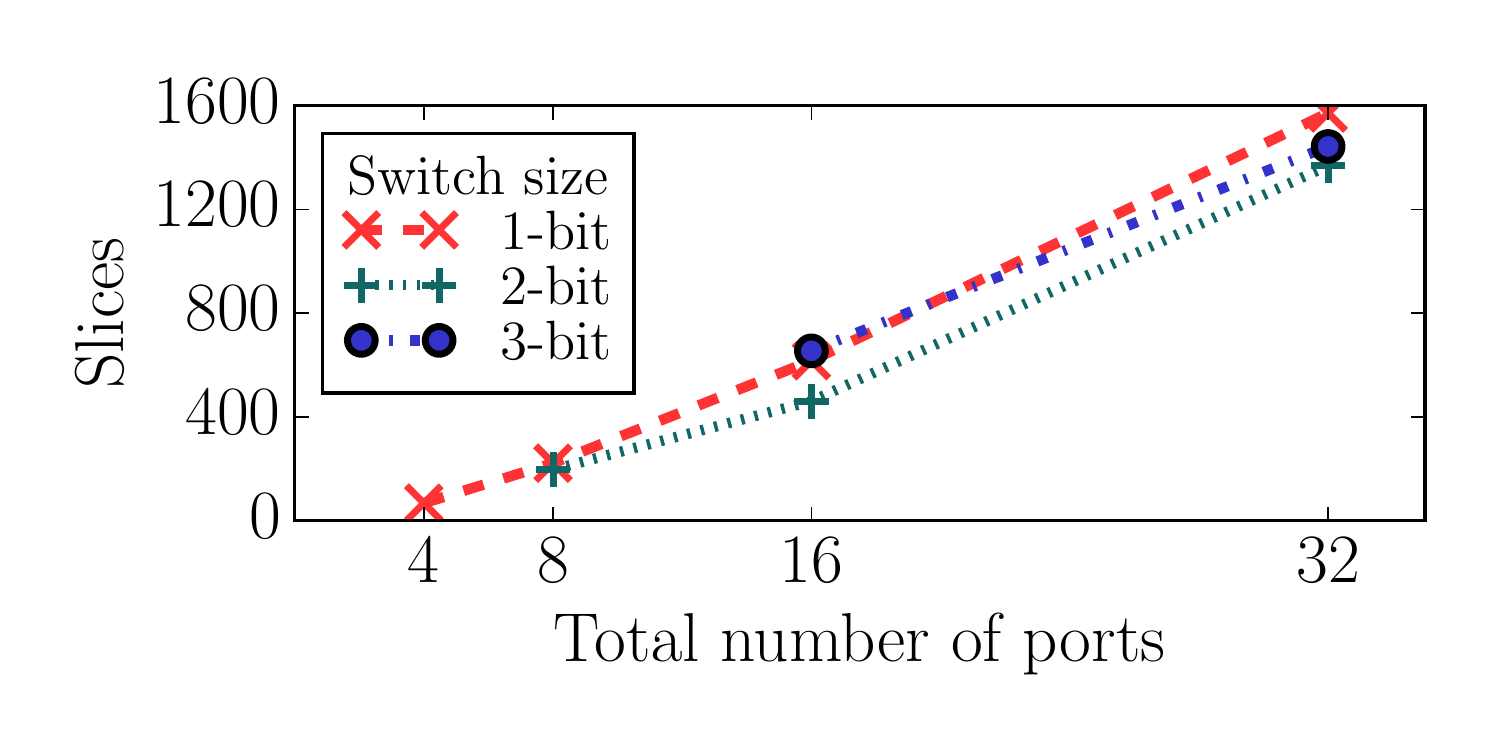}}\hfill
  \subfloat[Network bandwidth\label{fig:bw}]{\includegraphics[width=0.32\textwidth,clip,trim=0.5cm 0.5cm 0.5cm 0.5cm]{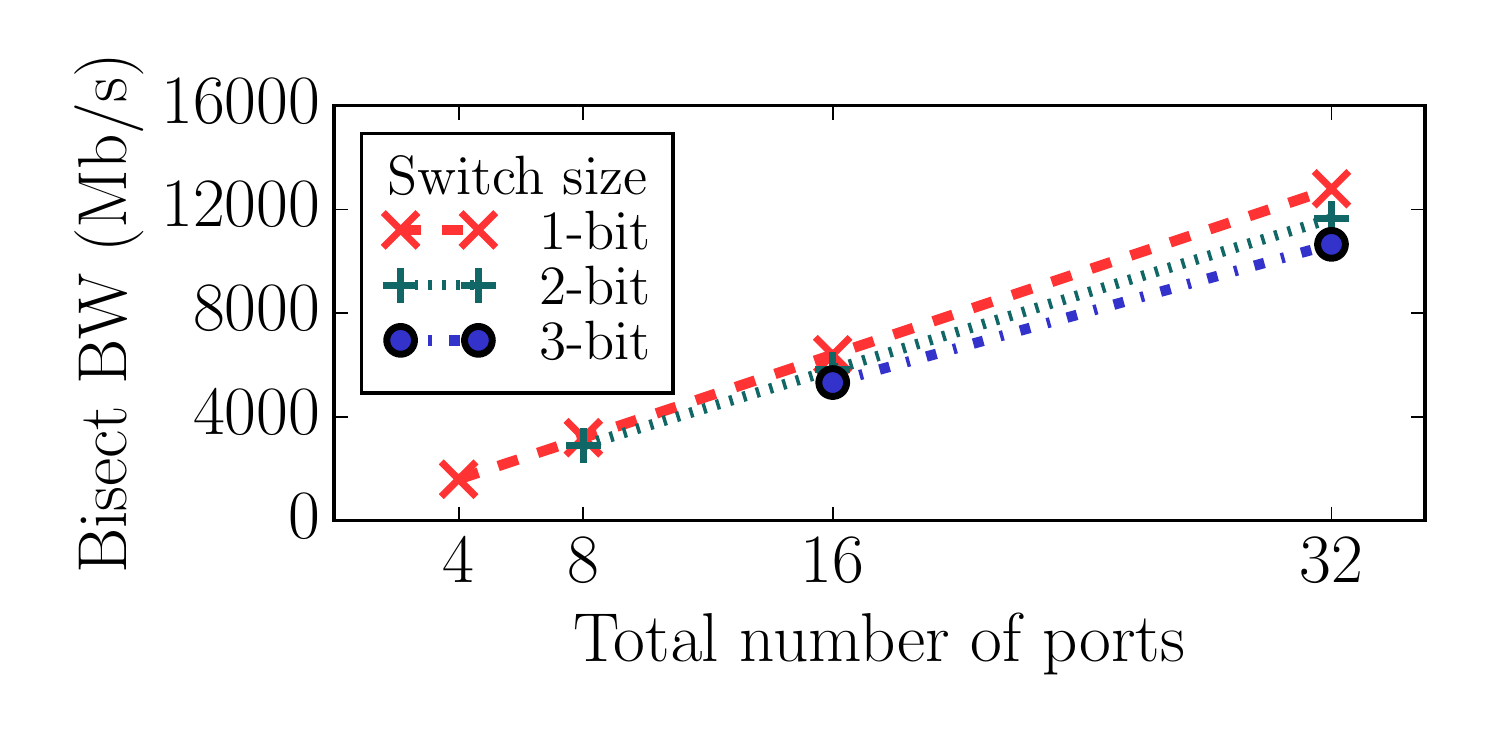}}\hfill
  \subfloat[Efficiency\label{fig:efficiency}]{\includegraphics[width=0.32\textwidth,clip,trim=0.5cm 0.5cm 0.5cm 0.5cm]{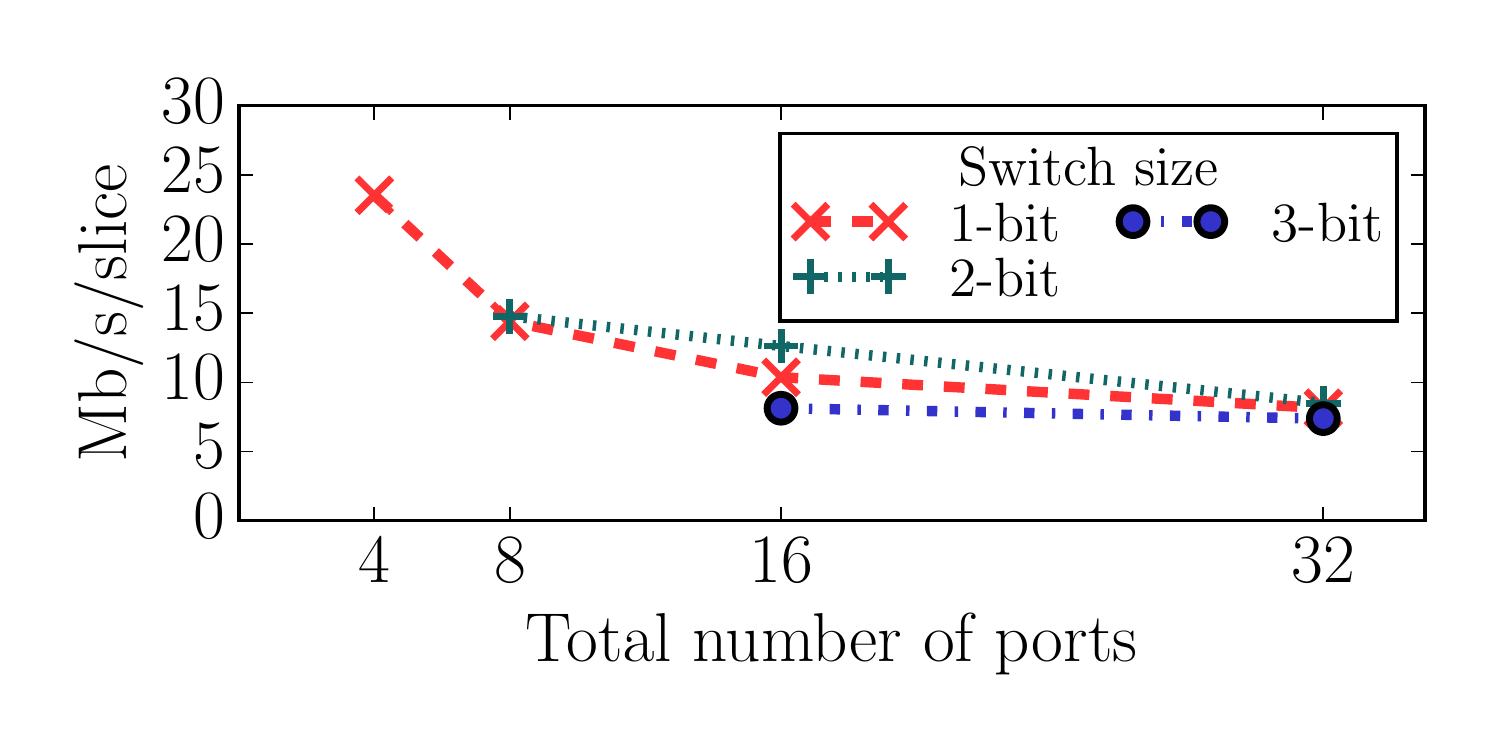}}
  \caption{Scaling behaviour of MCENoC when synthesized to FPGA target.}
  \label{fig:performance}
\end{figure*}

There is redundancy present in the routing configuration due to the in-band
control that is used. If a two-port switching element receives a route
configuration on one input, the other input could be implicitly routed to the
other available output. However, we choose to require the redundant
configuration, to make route setup time consistent across all inputs, as well
as to allow early identification of an erroneous route request. If implicit
routing was possible, an unexpected route will be identified in \emph{at least}
$2s$ cycles, whereas with redundant route configuration, it is
\emph{at most} $2p + s$ cycles.

Although no blocking takes place within the network, it can be imposed by the
connected nodes. The \texttt{cts} signal signifies that the source is clear to
send, and is asserted by default during route setup, allowing the route
configuration bits to be loaded into the network. Upon completing the route
setup, \texttt{cts} is then controlled by the destination node. If the
destination cannot receive more data, it must de-assert \texttt{cts}, and the
source must not transmit more data until it is re-asserted.

To accommodate transmission latency, the receiver should have a buffer of at
least $2s$ bits in size, and de-assert \texttt{cts} if less than this is
available. This prevents loss of data. If data production and consumption rates
can be determined statically, as may be the case for certain tasks in a
real-time context, then it may be possible to ignore the flow control signals,
as the switching element behaviour is not affected by \texttt{cts}.

\section{Scaling and performance}
\label{sec:evaluation}

In this section we illustrate the scaling properties of the MCENoC, examine
logic utilisation when various configurations are synthesised to FPGA, and
provide a comparison and evaluation against other designs.

The current design can operate at up to \SI{400}{\mega\hertz} according to the
Xilinx Vivado timing reporting tools, with the worst-case being
\SI{333}{\mega\hertz}. The size of the switching element governs
the achievable clock speed, rather than the network size, although a fully
integrated system may have other effects upon timing.

The architecture's performance and resource utilisation scaling are shown in
\cref{fig:performance}.  The FPGA slice utilisation (\cref{fig:util}) scales
superlinearly, but to a manageable degree for the network sizes testing.
Bisection bandwidth (\cref{fig:bw}) scales linearly with the number of nodes,
which is expected given that the operating frequency is fixed for each
switching element size. Consolidating these two metrics into a measure of
efficiency --- \SI{}{\mega\bit\per\second} per slice, shown in
\cref{fig:efficiency} --- indicates that the two-bit switching element is
marginally preferable over others for the network sizes that were synthesised.

To reason about integration with computational IP, we consider the Microblaze
core. Its device utilisation is dependent upon configuration, however an
example small configuration can use less than 800 slices\footnote{Evident
through public discussion: \url{http://tinyurl.com/j35p7ld}} in the Xilinx v7
architecture. This indicates that, conservatively speaking, a sixteen-core
Microblaze system and MCENoC network is possible on the Kintex-7K160T that we
are using as our current target, which has \SI{25}{\kilo\mbox{-slices}}
available.

The bisection bandwidth, $B$ of the MCENoC can be calculated through the simple
equation \cref{eq:bisection}, where $w$ is the bit-width of data transfer, $f$
is the operating frequency and $n$ is the number of ports. The current design
assumes each port uses 1-bit serialised data. Therefore, at
\SI{364}{\mega\hertz} (the operating frequency of our 2-bit switch
implementation), an eight-node MCENoC provides a bisection bandwidth of
\SI{2.9}{\giga\bit\per\second}. A 32-node system achieves in excess of
\SI{11.6}{\giga\bit\per\second}.
\begin{equation}
  B = f \times w \times n \quad \SI{}{\bit\per\second}.
  \label{eq:bisection}
\end{equation}

As an example comparison, the Epiphany E64G401\cite{Adapteva} has
\SI{102}{\giga\byte\per\second} bisection bandwidth in an ASIC that features
double the node count, 8-bit data width and three separate networks compared to
our MCENoC example. On a per-node, per-bit-width basis, the E64G401's network
achieves \SI{199}{\mega\bit\per\second} vs. the 32-node MCENoC's
\SI{725}{\mega\bit\per\second}. More in-depth evaluation would be better served
by application-specific case studies.

\section{Routing and scheduling}
\label{sec:scheduling}

It is proven in\cite{Benes1962} that a route exists for any permutation of
$1:1$ communication between all nodes in a Bene\u{s} network. However,
calculating the routes for a permutation without global knowledge is
non-trivial\cite{Pan2008}. For the MCENoC, we assume that for pre-defined
tasks, routes are determined statically, thus online methods are not required.
There are still design decisions to make, in the form of switch size selection,
as well as scheduling strategies. This section investigates the implications of
the network design with respect to these issues.

\subsection{Network equivalence}

\begin{figure}
  \centering
  \subfloat[Two-port switching elements.\label{fig:twoport-route}]{
    \ifarxiv
      \centering
      \includegraphics[width=0.8\columnwidth]{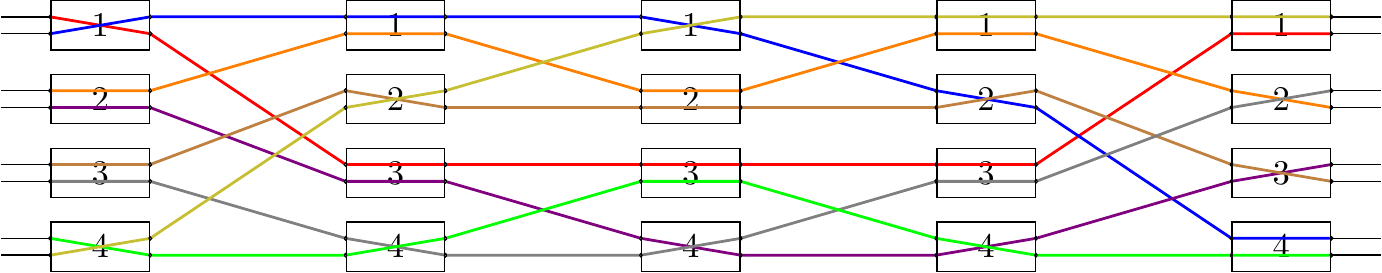}
    \else
      \begin{tikzpicture}[scale=0.35,every node/.style={scale=0.35}]
        \input{media/net-32port-1bit.tikz}
      \end{tikzpicture}
    \fi
  }\\
  \subfloat[Four-port and two-port elements.\label{fig:fourport-route}]{
    \ifarxiv
      \centering
      \includegraphics[width=0.8\columnwidth]{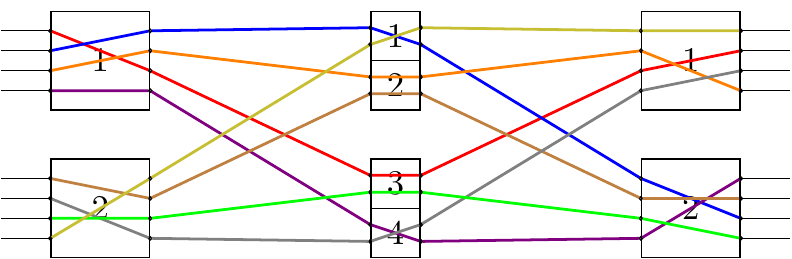}
    \else
      \begin{tikzpicture}[scale=0.75,every node/.style={scale=0.6}]
        \input{media/net-8port-2bit.tikz}
      \end{tikzpicture}
    \fi
  }
  \caption{Route equivalence in two eight-node MCENoC implementations.}
  \label{fig:route-example}
\end{figure}

Routing algorithms such as \cite{Waksman1968} assume two-port switches as per a
traditional Bene\u{s} network, where our header-based configuration requires
one bit per switch. However, networks with higher degree switching elements can
be equivalent, simply consuming a larger number of header bits.
\Cref{fig:route-example} shows an example 8-port permutation in two network
variants. Taking a single route from this --- connecting port 0 to port 1 ---
the header bits are \texttt{10001}. In \cref{fig:twoport-route}, a single bit
is used, left to right, to configure each stage. In \cref{fig:fourport-route},
two bits are used in the outer stages and a single bit in the middle stage,
interpreted as \texttt{10-0-01}, giving the same connection.

\subsection{Scheduling}

In a practical system, a single routing permutation will be insufficient, as
each node may have several other nodes with which it needs to communicate. This
can be resolved through Time Division Multiplexing (TDM) of permutations. We
assume that the nodes are responsible for ensuring the time divisions are
adhered to, keeping the switching elements and network control simple, but
requiring tightly synchronised timing between nodes. Such synchronisation is
tractable due to the predictable latency of the network allowing precise time
exchange between nodes. This can take place periodically, or be embedded into
each permutation.

A Bene\u{s} style network combined with TDM abstracts other topologies well.
For example, to emulate a 2D mesh, the MCENoC network can be divided into four
permutations, with each permutation representing one of four directions. Adding
dimensions requires two additional permutations. A broadcast from a single node
can reach all other nodes in $\log_2(n + 1)$ permutations as the number of
nodes that can forward the message doubles with each permutation.

\begin{figure}
  \centering
  \includegraphics[width=1.0\columnwidth,clip,trim=0.5cm 0.5cm 0.5cm 0.5cm]{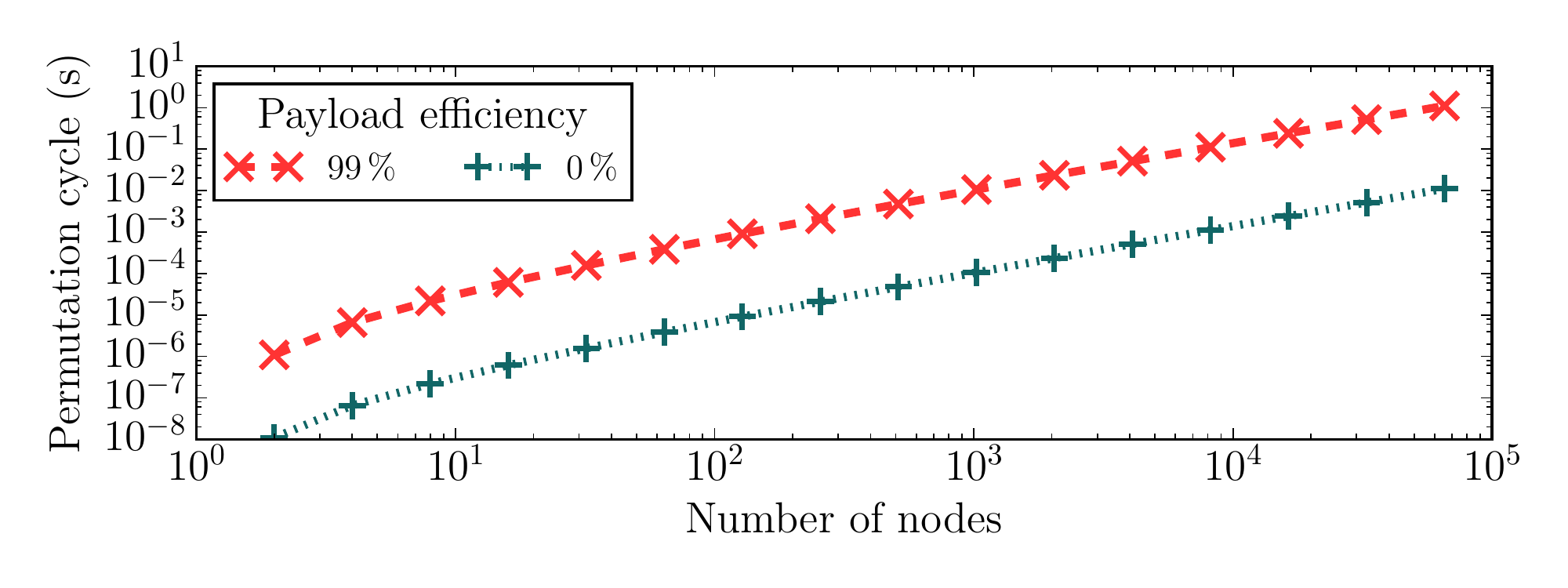}
  \caption{Time taken to cycle through N communications in an N-node network at
    \SI{364}{\mega\bit\per\second\per{port}}, considering a desired payload
  efficiency.}
  \label{fig:tdmscale}
\end{figure}

Applying TDM to the network increases the route calculation time linear to the
number of permutations required. The low overhead of the in-band route setup
enables low latencies, even between instances of a particular permutation where
a large number of permutations are used. The duration of each permutation can
be adjusted depending on the payloads that need to be transmitted. An example
performance profile of a worst-case TDM schedule in which all nodes talk
directly to all other nodes is shown in \cref{fig:tdmscale}.

This assumes $N$ TDM phases are needed, are of equal duration, and have a set
efficiency achieve by sizing the payload appropriately in relation to the route
setup and tear-down latency. Applying this conservative model gives a worst
case time between cycles of the TDM of \SI{914}{\micro\second} for 128 cores
with \SI{99}{\percent} payload efficiency. In a \SI{65}{\kilo\mbox{-node}}
system this reaches \SI{1.12}{\second}. However, the time per message is
\SI{17.03}{\micro\second}. Therefore, with repetition of routes within the TDM
cycle, the latency between critical communications can be kept very low.

The underlying switching architecture is sufficiently flexible and predictable
that scheduling even complex communication patterns is tractable. Low latency
and predictability is favoured over high bandwidth, although the network
utilisation can be kept close to \SI{100}{\percent} if the number of
permutations and arrangement of communications is carefully selected.

To avoid the need for all traffic to be statically scheduled, a dynamic system
can be used during pre-allocated slack phases in the network.
In\cite{KostrzewaScheduling2016}, a resource controller grants network time to
requesting nodes, managing available slack in the system's mesh network. For
strict and unimpeded control, separate control and data networks can be used. A
similar approach can also be applied to the MCENoC, using TDM to provide this
separation.

\section{Formal verification}
\label{sec:formal}

\begin{figure*}
  \vspace{-1.5em}
  \centering
  \subfloat[Num switches\label{fig:numsw}]{\includegraphics[width=0.32\textwidth,clip,trim=0.5cm 0.5cm 0.5cm 0.5cm]{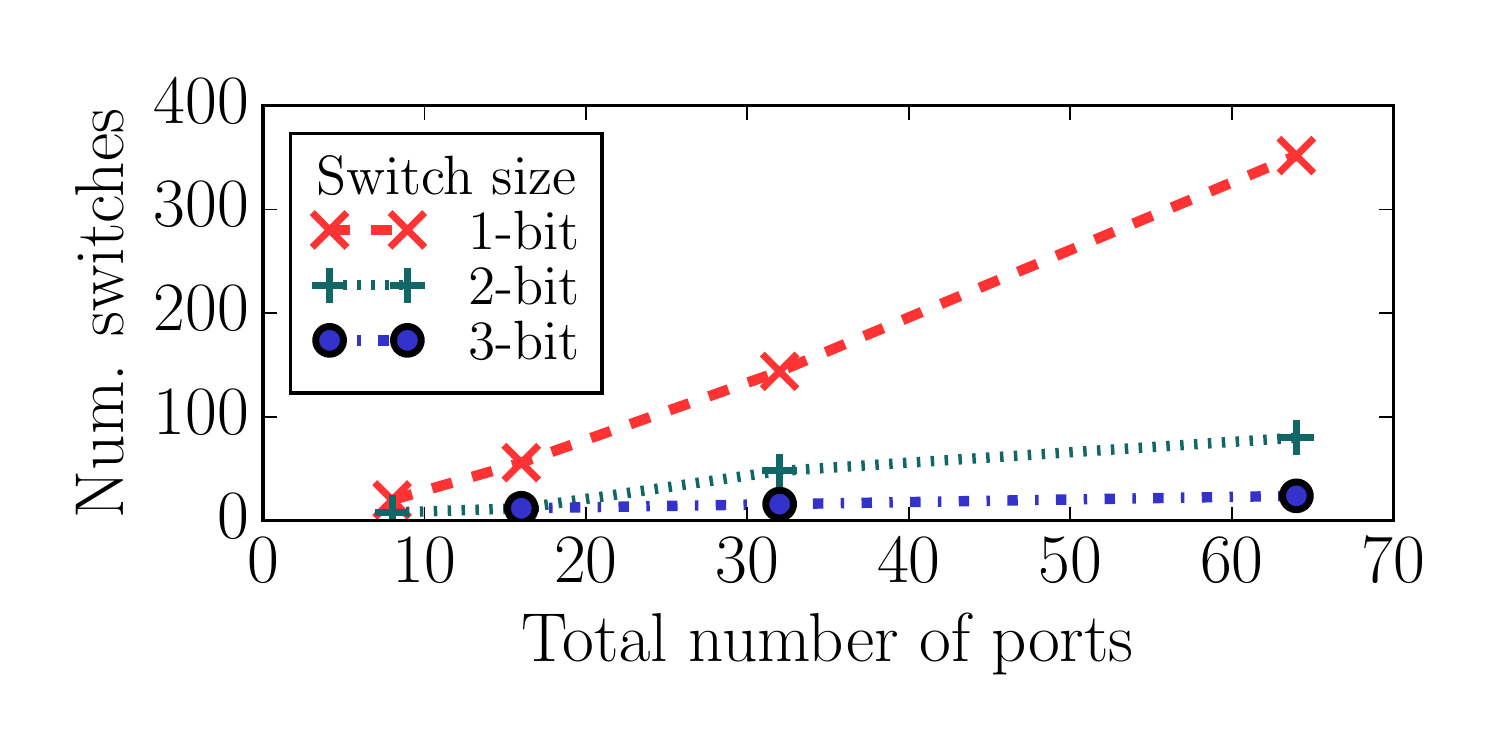}}\hfill
  \subfloat[Core proof time\label{fig:coretime}]{\includegraphics[width=0.32\textwidth,clip,trim=0.5cm 0.5cm 0.5cm 0.5cm]{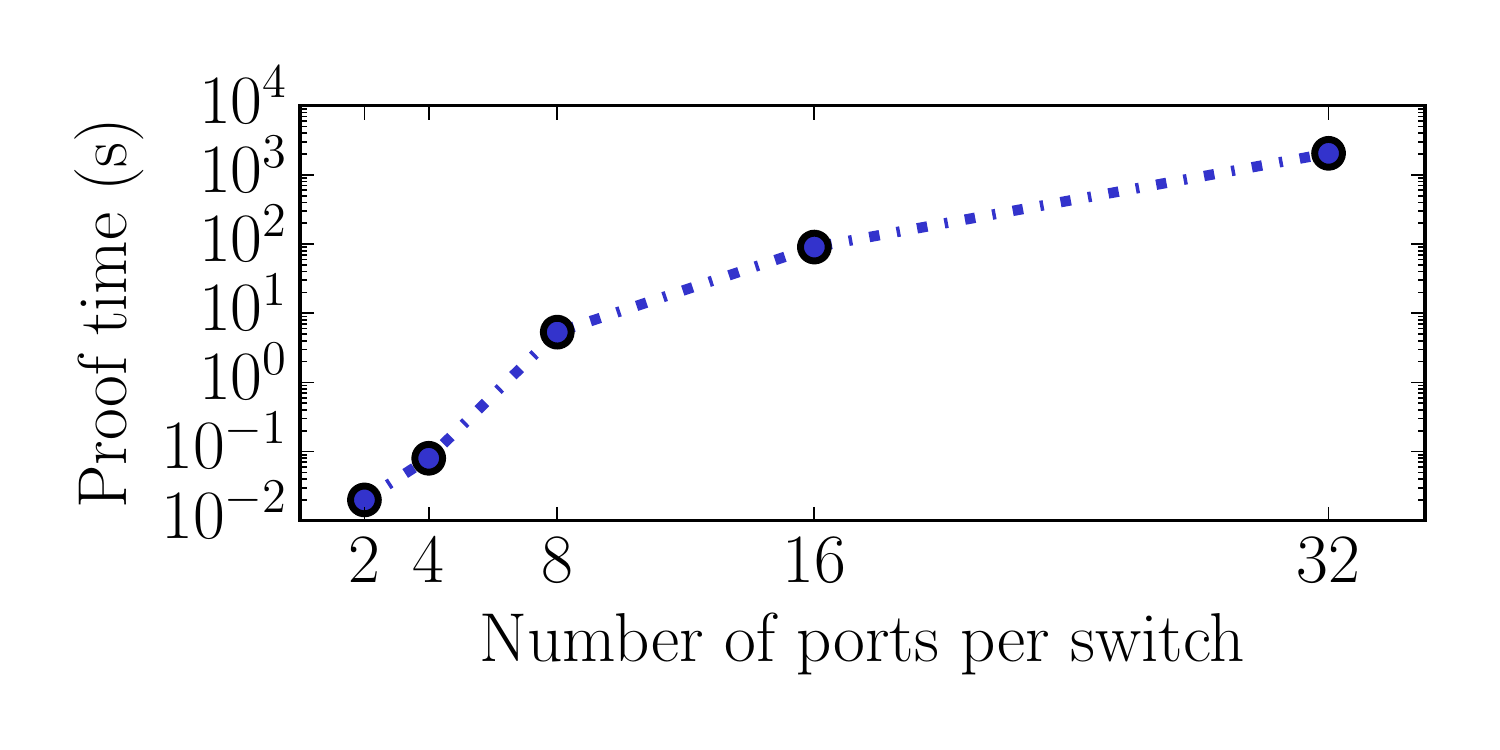}}\hfill
  \subfloat[Network proof time\label{fig:nettime}]{\includegraphics[width=0.32\textwidth,clip,trim=0.5cm 0.5cm 0.5cm 0.5cm]{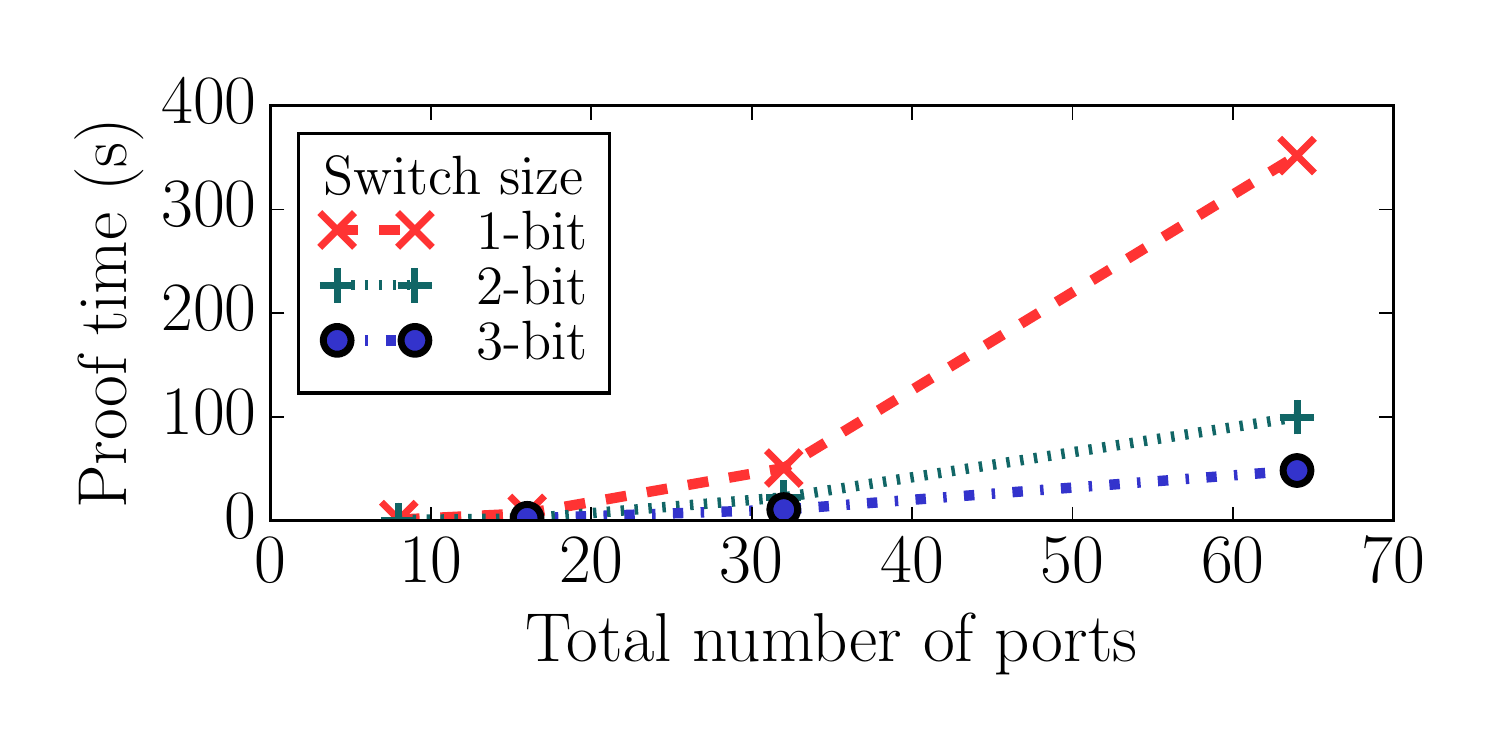}}
  \caption{Performance and scaling of the MCENoC formal verification}
  \label{fig:proofperformance}
\end{figure*}

Formal methods of verification\cite{seligman2015formal} allow properties of a
design to be specified and proven exhaustively, covering all parts of the
design's state-space that could be affected by that particular property. This
is achieved with the use of formal verification tools (in the case of this
paper, JasperGold), which can prove or disprove these properties, providing
counterexample traces in the case of failure. This has benefits over
traditional, test-driven verification, in that it does not rely on random or
pre-defined test vectors to expose design flaws or implementation faults.

Complete verification coverage requires that properties describe all aspects of
the design. Further, the exhaustive nature of formal verification means that complex properties and behaviours can lead to infeasibly long proof times.

The construction of formal properties also serves as a test of the
specification, requiring it to be formalised in a syntax more rigid than
natural language. Thus, properties can expose deficiencies in an
implementation, the way in which the specification is defined, as well as its
interpretation.

In formal verification terminology, a \emph{property} is a formal description
of the behaviour of a device over a period of time. This is typically
formulated as a sequence of preconditions, which if met, a postcondition
sequence must occur. If a property is \emph{asserted}, then a verification
environment (formal or simulation-based) will trigger a failure if the property
is ever found not to hold. If a property is \emph{assumed}, then it constrains
the verification to consider only states where it holds true.

The MCENoC is designed with formal verification as the intended verification
method. Given that is is constructed from simple, replicated elements, a large
portion of the proof process is possible within seconds. In the rest of this
section we examine the steps necessary for formal proof of MCENoC, as well as
the performance scaling of the proof process at various system levels and
sizes.

\subsection{Specification}

To form the specification of the design, we take the general requirements from \cref{sec:requirements} and produce more specific criteria. These are classified at several levels:

\begin{description}[font=\normalfont\itshape]

  \item[C: Core switching elment level.] The behaviour of a
    single switch and its interfaces.

  \item[N: Network level.] Connectivity to the edge of the Bene\u{s}
    network, signal propagation and routing behaviour within.

  \item[S: System level.] Behaviours when interacting with nodes.

\end{description}

Example specification criteria are given in \cref{tab:spec}. At the core and
network level, all specification items need to have one or more properties that
fully describe the relevant behaviour of the MCENoC. However, the system level
includes considerations outside of the scope of the MCENoC itself. It is still
possible to define properties at the system level, but it may not be possible
to verify them formally when solely examining the MCENoC implementation.
However, these properties can still be used as assumptions that assist in the
proof of other, lower level specification items.

\begin{table}
  \centering
  \caption{Examples of specification criteria at each design level.}
  \label{tab:spec}
  \begin{tabular}{cp{0.5\columnwidth}l}

    \textbf{ID} & \textbf{Description} & \textbf{Property} \\ \hline

    C1 & No two active inputs can share the same output channel. & \scriptsize
    \texttt{no\_shared\_direction} \\ 

    C15 & A port rejects further data in the event of an inbound error signal
    and propagates the error signal. & \scriptsize \texttt{reject\_on\_err} \\

    N4 & All routes through the network connect to the expected destination
    if no routing conflict arises and the target port does not assert an error.
    & \scriptsize \texttt{route\_correct} \\ 

    S4 & Higher priority tasks must create their routes before lower priority
    tasks. & N/A \\ \hline
  \end{tabular}
  \vspace{-5mm}
\end{table}

\subsection{Property definition}

Properties are defined in SVA\cite{Mehta2014}, and then asserted for each of
the ports or instances defined in any configuration of the system. Checking of
these properties is performed using the Cadence JasperGold formal verification
tool.

\Cref{lst:c1prop} gives an example property for a core switching element,
seeking to prove that specification criteria C15 is met.  This criteria refers
to the rejection of further data if an error signal is received, and that the
error signal propagates. Thus, C15 ensures that on error, a switch stops
forwarding on the connected port, but communicates the error backwards through
the established route.

The property is defined as \texttt{reject\_on\_err(i)} where \texttt{i} is the
port number. The portion before the \texttt{|=>} symbol signifies the
pre-condition, stating that the port is connected and that the connected error
signal has risen in the current clock cycle. Checking of the post-condition
begins one cycle later. This is a two-cycle sequence, where first the port is
expected to enter the \texttt{ABORT} state, which should include the error
output being asserted downstream. After another cycle (signified by
\texttt{\#\#1}), the port's forward connections should all be de-asserted,
releasing the forward connection.

\vspace{-0.25em}
\begin{algorithm}
  \caption{Property example to check criterion C15}
  \label{lst:c1prop}
  \begin{lstlisting}
property reject_on_err(i);
    portstate[i] == ACCEPT and
    $rose(ports.err_in[direction[i]])
  |=>
    portstate[i] == ABORT and ports.err_out[i]
    ##1 !(ports.clm_out[direction[i]] |
      ports.act_out[direction[i]] |
      ports.dat_out[direction[i]] );
endproperty
  \end{lstlisting}
\end{algorithm}
\vspace{-0.5em}

This property can only fail if the pre-condition is met and then the
post-condition sequence contradicts the definition. A formal verification tool
can provide coverage in both respects, first demonstrating that the
pre-condition is reachable, and then proving exhaustively that, under any
condition where the pre-condition is reached, the post-condition sequence holds
true.

\subsection{Proof performance}

The performance of the verification process was conducted on a dual-socket Intel Xeon X5460 server. Default JasperGold prover settings were used. Two levels of verification were tested: a single switching element and a full network. Results are shown in \cref{fig:proofperformance}.

\Cref{fig:numsw} demonstrates the scaling of the number of switches with
network size, depending on the size of each switching element.
\Cref{fig:coretime} shows the proof time for core switching element properties.
The proof time grows exponentially with the port count, which is an expected
outcome as the number of generated assertions increases, as does the internal
state of the switch. This demonstrates that the proof effort is tractable for
sizes that are desirable for synthesis.

In \cref{fig:nettime}, the full network proof time is demonstrated. This
includes network level properties as well as re-proof of switch-level
properties. The majority of the proof burden is dependent upon the number of
switches, hence using higher order switches reduces the overall proof time.
However, these results omit the checking of \texttt{route\_correct}
(specification N4 in \cref{tab:spec}), which on its own requires approximately
90~minutes to prove on an 8-port network of 1-bit switches. This is due to the
length of the pre-condition sequence it uses, which in many cases does not
reach completion before some other correct behaviour (such as a routing
conflict) takes place. To overcome this, one can rely on the switch-level
properties combined with checking of the connectivity, which is a static
property, to achieve the same guarantee. Alternatively, properties defining
specific routing schedules could be defined, for example constrained to a
particular application definition, in order to provide a smaller search space
to the prover.

In summary, our performance evaluation demonstrates that formal proof of the MCENoC design that is at a scale practical for realisation in hardware, is tractable and timely using current tools. Potential performance issues have been exposed, and mitigation methods recommended.

\section{Conclusions}
\label{sec:conclusions}

A novel NoC implementation has been presented for multi-core MCE systems that
addresses the problem of predictability in systems of growing scale and
complexity. In answer to the requirements we define for such systems, this work
uses Bene\u{s}-type network structures to achieve equidistant communication
between nodes that remains scalable, providing an implementation that can be
in-band controlled and has clearly defined route creation priorities. The
network structure and switch behaviour can be combined with static scheduling
and time division multiplexing to provide predictable low-latency communication
that guarantees deadlines will not be missed and that critical communications
are not interfered with, aiding software-level timing analysis and system
certification.

The design is formally verified, providing exhaustive proof that specified
behaviours hold, ensuring both that the specification is well defined and that
the implementation is correct. Current proof tools can verify designs of sizes
that are practical for synthesis to FPGA, whereupon current synthesis results
indicate a typical \SI{364}{\mega\hertz} operation and a bisection bandwidth of
\SI{11.6}{\giga\bit\per\second} for an $32 \times 32$ size network.

\subsection*{Future work}

Network nodes, in the form of memories, processing and peripherals, may be
integrated with the design to demonstrate it at the system level. This can
include synthesis to FPGA as well as fully-simulated, potentially relying on
the proofs made to simplify the network simulation process. The data-path of
the design can also be expanded from 1-bit in order to explore the trade-off
between bandwidth and resource utilisation, including FPGA slices and I/O.

Proof of properties at the system level, where node and software behaviour must
be considered, is an important next step in exploiting the predictable nature
of the MCENoC. This may extend beyond the SystemVerilog of the MCENoC, however,
requiring innovative proof techniques.

At the software and tool-chain levels, the implementation of routing algorithms
for required communication patterns is the subject of ongoing work. Routing
must be reconciled with task priorities and network utilisation through time
division multiplexing as discussed in \cref{sec:scheduling}, which can be
evaluated through MCE application case studies.

\section*{Acknowledgment}

\ifarxiv

The research leading to these results has received funding from the ARTEMIS
Joint Undertaking under grant agreement number 621429 (project EMC2).

\else

  Acknowledgements currently blank for review.

\fi
\bibliographystyle{IEEEtran}
\bibliography{EMC2,swallow-refs}  %

\begin{thebibliography}{10}
\providecommand{\url}[1]{#1}
\csname url@samestyle\endcsname
\providecommand{\newblock}{\relax}
\providecommand{\bibinfo}[2]{#2}
\providecommand{\BIBentrySTDinterwordspacing}{\spaceskip=0pt\relax}
\providecommand{\BIBentryALTinterwordstretchfactor}{4}
\providecommand{\BIBentryALTinterwordspacing}{\spaceskip=\fontdimen2\font plus
\BIBentryALTinterwordstretchfactor\fontdimen3\font minus
  \fontdimen4\font\relax}
\providecommand{\BIBforeignlanguage}[2]{{%
\expandafter\ifx\csname l@#1\endcsname\relax
\typeout{** WARNING: IEEEtran.bst: No hyphenation pattern has been}%
\typeout{** loaded for the language `#1'. Using the pattern for}%
\typeout{** the default language instead.}%
\else
\language=\csname l@#1\endcsname
\fi
#2}}
\providecommand{\BIBdecl}{\relax}
\BIBdecl

\bibitem{itrs2013}
A.~B. Kahng, ``{The ITRS design technology and system drivers roadmap},'' in
  \emph{Proceedings of the 50th Annual Design Automation Conference on - DAC
  '13}.\hskip 1em plus 0.5em minus 0.4em\relax New York, New York, USA: ACM
  Press, 2013.

\bibitem{mcavionics2012}
J.~Nowotsch and M.~Paulitsch, ``{Leveraging Multi-core Computing Architectures
  in Avionics},'' in \emph{2012 Ninth European Dependable Computing
  Conference}.\hskip 1em plus 0.5em minus 0.4em\relax IEEE, may 2012, pp.
  132--143.

\bibitem{tterationale}
H.~Kopetz, ``{The Rationale for Time-Triggered Ethernet},'' in \emph{2008
  Real-Time Systems Symposium}.\hskip 1em plus 0.5em minus 0.4em\relax IEEE,
  nov 2008, pp. 3--11.

\bibitem{cacheCoherency}
M.~M.~K. Martin, M.~D. Hill \emph{et~al.}, ``{Why on-chip cache coherence is
  here to stay},'' \emph{Communications of the ACM}, vol.~55, p.~78, jul 2012.

\bibitem{mceCert2010}
S.~Baruah, H.~Li \emph{et~al.}, ``{Towards the Design of Certifiable
  Mixed-criticality Systems},'' in \emph{2010 16th IEEE Real-Time and Embedded
  Technology and Applications Symposium}.\hskip 1em plus 0.5em minus
  0.4em\relax IEEE, apr 2010, pp. 13--22.

\bibitem{DBLP:journals/tecs/WilhelmEEHTWBFHMMPPSS08}
R.~Wilhelm, J.~Engblom \emph{et~al.}, ``{The worst-case execution-time problem
  - overview of methods and survey of tools},'' \emph{ACM Trans. Embedded
  Comput. Syst.}, vol.~7, 2008.

\bibitem{XeonPhi2013}
{Intel Corporation}, ``{Intel Xeon Phi Coprocessor},'' Intel Corporation, Tech.
  Rep., 2013.

\bibitem{n3xt}
M.~M. {Sabry Aly}, M.~Gao \emph{et~al.}, ``{Energy-Efficient Abundant-Data
  Computing: The N3XT 1,000x},'' \emph{Computer}, vol.~48, pp. 24--33, dec
  2015.

\bibitem{Bell2008}
S.~Bell, B.~Edwards \emph{et~al.}, ``{TILE64 processor: A 64-core SoC with mesh
  interconnect},'' in \emph{Digest of Technical Papers - IEEE International
  Solid-State Circuits Conference}, vol.~51, 2008.

\bibitem{Adapteva}
Adapteva, ``{E64G401 EPIPHANY 64-core microprocessor datasheet}.''

\bibitem{PicoChip}
A.~Duller, G.~Panesar \emph{et~al.}, ``{Parallel Processing the picoChip
  way},'' \emph{Communicating Processing Architectures}, pp. 299--312, 2003.

\bibitem{REPAIR}
R.~Abdel-khalek and V.~Bertacco, ``{Correct Runtime Operation for NoCs through
  Adaptive-Region Protection},'' in \emph{Design, Automation Test in Europe
  Conference Exhibition (DATE), 2016}, 2016.

\bibitem{3dredundancy}
M.~Ebrahimi, M.~Daneshtalab \emph{et~al.}, ``{Fault-tolerant method with
  distributed monitoring and management technique for 3D stacked meshes},'' in
  \emph{The 17th CSI International Symposium on Computer Architecture {\&}
  Digital Systems (CADS 2013)}.\hskip 1em plus 0.5em minus 0.4em\relax IEEE,
  oct 2013, pp. 93--98.

\bibitem{Clos1952}
C.~Clos, ``{A Study of Non-Blocking Switching Networks},'' \emph{Bell System
  Technical Journal}, pp. 406--424, 1952.

\bibitem{Benes1962}
V.~E. Bene{\v{s}}, ``{On Rearrangeable Three-Stage Connecting Networks},''
  \emph{Bell System Technical Journal}, vol.~41, pp. 1481--1492, sep 1962.

\bibitem{Pan2008}
C.~Scheideler, \emph{{Universal Routing Strategies for Interconnection
  Networks}}, ser. Lecture Notes in Computer Science.\hskip 1em plus 0.5em
  minus 0.4em\relax Berlin, Heidelberg: Springer Berlin Heidelberg, 1998, vol.
  1390.

\bibitem{jiang2014}
Y.~Jiang and M.~Yang, ``{On circuit design of on-chip non-blocking
  interconnection networks},'' in \emph{2014 27th IEEE International
  System-on-Chip Conference (SOCC)}, vol.~40, no.~8.\hskip 1em plus 0.5em minus
  0.4em\relax IEEE, sep 2014, pp. 192--197.

\bibitem{Mehta2014}
A.~B. Mehta, \emph{{SystemVerilog Assertions and Functional Coverage: Guide to
  Language, Methodology and Applications}}.\hskip 1em plus 0.5em minus
  0.4em\relax New York, NY: Springer New York, 2014, ch. System Verilog
  Assertions, pp. 9--28.

\bibitem{Waksman1968}
A.~Waksman, ``{A Permutation Network},'' \emph{Journal of the ACM}, vol.~15,
  pp. 159--163, jan 1968.

\bibitem{KostrzewaScheduling2016}
A.~Kostrzewa, S.~Saidi \emph{et~al.}, ``{Slack-Based Resource Arbitration for
  Real-Time},'' in \emph{Design, Automation Test in Europe Conference
  Exhibition (DATE), 2016}, 2016, pp. 1012--1017.

\bibitem{seligman2015formal}
E.~Seligman, T.~Schubert \emph{et~al.}, \emph{Formal Verification: An Essential
  Toolkit for Modern VLSI Design}.\hskip 1em plus 0.5em minus 0.4em\relax
  Morgan Kaufmann, 2015.

\end{thebibliography}

\end{document}